\documentclass[aps,prl,twocolumn,reprint,amsmath,amssymb]{revtex4-2}
\usepackage{graphicx}
\usepackage{dcolumn}
\usepackage{amsthm}
\usepackage{bm}
\usepackage{bbm}
\usepackage{mathrsfs}
\usepackage{xcolor}
\newcommand{\red}[1]{\textcolor{red}{#1}}

\usepackage[colorlinks=true, pdfstartview=FitV, linkcolor=blue, citecolor=blue, urlcolor=blue]{hyperref}
\usepackage{nameref}
\begin{document}

\title{Magnon Thermal Hall Effect Induced By Symmetric Exchange Interaction}

\author{Jikun Zhou}

\affiliation{CAS Key Laboratory of Strongly-Coupled Quantum Matter Physics and Department of Physics, University of Science and Technology of China, Hefei, Anhui 230026, China
}

\author{Yang Gao}
\email[Correspondence author:~~]{ygao87@ustc.edu.cn}
\affiliation{CAS Key Laboratory of Strongly-Coupled Quantum Matter Physics and Department of Physics, University of Science and Technology of China, Hefei, Anhui 230026, China
}
\affiliation{Hefei National Laboratory, University of Science and Technology of China, Hefei 230088, China}

\author{Qian Niu}
\affiliation{CAS Key Laboratory of Strongly-Coupled Quantum Matter Physics and Department of Physics, University of Science and Technology of China, Hefei, Anhui 230026, China
}
\date{\today}

\begin{abstract}
By analyzing the spin-group symmetry of magnons, we establish two generalized Onsager's relations in the magnon thermal Hall effect, which reveals the rich and complicated structures of the magnon Berry curvature in the parameter space of different types of the exchange coupling.  As an important consequence, we find that the diagonal and off-diagonal part of the symmetric anisotropic exchange coupling together can support the planar magnon thermal Hall effect without the DM interaction. This removes the local intersion-symmetry-breaking condition for generating the magnon thermal Hall effect. Moreover, we predict an exotic phenomenon that the thermal conductivity exhibits angular dependence with respect to the in-plane magnetization.  Our work lays the ground for decoding the coupling between magnon transport and different types of exchange interactions.
\end{abstract}

\maketitle

{\it Introduction.}---As potential carriers in energy and spin transport, magnons, elementary excitations of spin orders, have drawn much attention in recent years. One outstanding example is the magnon thermal Hall effect\cite{Onose2010}, which generates a thermal current perpendicular to the applied temperature gradient. The corresponding thermal conductivity can be related to the Berry curvature of magnon in the momentum space~\cite{Matsumoto2011prb, Matsumoto2011prl, Katsura2010, Matsumoto2014}. As a fundamental rule, the Onsager's reciprocal relation\cite{onsager1,onsager2} restricts the structure of the transport coefficient. In the electronic counterpart, i.e., the anomalous Hall effect\cite{NagaosaAnomalous}, it enforces the corresponding conductivity to be an odd function of the spin orientation, reveals the intimate relation between the band geometry and the order parameter, and hence serves as a guiding principle for experimental detection. In the magnonic case, however, besides separate case studies, a comprehensive understanding of the Onsager's relation is still missing.

The difficulty lies in the identification of the critical parameter in the magnon transport. It is widely believed that the Dzyaloshinskii-Moriya~(DM) interaction\cite{DZYALOSHINSKY1958241,Moriya1960} is indispensable in the magnon thermal Hall effect, associated with the breaking of the effective time-reversal symmetry consisting of a spin rotation and real time-reversal operation~\cite{Chen2014, Suzuki2017, Mook2019,Cheng2016,Zhang2019}, so that the thermal Hall conductivity is an odd function of the DM interaction. \red{Magnon thermal Hall effect induced by DM interaction has been studied in various systems theoretically and experimentally, such as honeycomb lattice\cite{Owerre2016, Owerre2016prb, Li2021}, kagome lattice\cite{Zhuo2021,Mook2014, Mook2019, Laurell2018, Hirschberger2015prl, Akazawa2020, Ideue2012}, square lattice\cite{BUzo2024,Kawano2019} and Lieb lattice\cite{Cao2015}.} However, it has been long overlooked that due to the spin-group symmetry, the choice of the spin rotation axis contains large gauge freedom, which essentially mixes the isotropic, anisotropic and antisymmetric part of the exchanging coupling. This complexity has been implicitly envisioned in recent studies~\cite{Neumann2022} which show that the anisotropic exchange coupling~\cite{Shekhtman1992,wan2023} which comes from second order perturbation of spin-orbit coupling may also induce a nontrivial magnon thermal Hall effect. To accurately identify the critical parameter in the magnon thermal Hall effect, a detailed analysis of the effective time reversal symmetry from the spin-group perspective is required. 

In this work, we provide such a theory and derive two generalized Onsager's relations, by identifying the group $\tilde{G}$ of full effective time reversal operations~(see Eq.~\eqref{eq_ftr}), inherited from the spin-only group\cite{Liu2022,spin-group2,spin-group3,spin-group4,spin-group5,spin-group6, Chen2025,Corticelli2022} in the original magnetic lattice, and recognizing two types of effective time reversal operations. They reveal the rich and complicated dependence of the magnon Berry curvature on various exchange couplings. As an important consequence, we establish the necessary condition when the anisotropic interaction alone can induce magnon thermal Hall without the DM interaction. \red{Since the anisotropic exchange coupling does not need to break the local inversion symmetry, our theory then expands potential systems with the magnon thermal Hall effect. Examples are ${\rm VAu_4}$ and monolayer ${\rm CrCl_3}$. Moreover, we show that the anisotropic exchange coupling can lead to an exotic in-plane magnon thermal Hall effect where the thermal Hall conductivity depends on in-plane orientation of magnetization.} 

{\it Local formulation of magnons.}---We first set up the general framework for discussing the effective time reversal operation and its implication. We consider periodic spin textures and assume that there can be arbitrary number of spins in a unit cell, each of which can point along different directions. Instead of choosing a global coordinate system for the spin texture, we adopt the following rule: we define local coordinate systems on different lattice sites such that the $z$-axis of the local frame always points along the equilibrium direction of the local spin order. Such local coordinate frame defined in this way is periodic. 

We now write down the general spin Hamiltonian within this local frame. As the topic in concern is the transport of the spin wave, we focus on the Hamiltonian of the fluctuation around the equilibrium spin texture and ignore how such texture is achieved. For a stable spin texture, it is required that the energy is quadratic and positive with respect to any fluctuation. With this in mind, the spin Hamiltonian should take the following quadratic form:
\begin{equation}\label{eq_bilinearH}
    H=\frac{1}{2}\sum_{n,m}{\sum_{i,j}^N \sum_{ab}{\boldsymbol{S}_{ni}^{a}J^{ab}_{ni,mj}\boldsymbol{S}^b_{mj}}},
\end{equation}
where $S_{ni}^a$ is the $a$-th component of the local spin on the $i$-th lattice site in the $n$-th unit cell, and $J^{ab}_{ni,mj}$ contains all exchange interactions between local spins. Moreover, $J^{ab}_{ni,mj}=0$ when one of the $ab$ indices takes $z$ and the other one takes $x$ or $y$. For definiteness, we assume that there are $N$ different spins in a unit cell. It is clear that Eq.~\eqref{eq_bilinearH} contains all the quadratic forms of the fluctuation. 

We now put the spin Hamiltonian into the canonical form. Through the Fourier transform and the Holstein-Primakoff transformation~\cite{Holstein1940}, the spin Hamiltonian can be expressed in the momentum space using the Nambu basis up to a constant: $H=\sum_{\bm{k}} \psi_{\bm{k}}^\dagger \mathcal{H}_{\bm{k}} \psi_{\bm{k}}$ with
\begin{equation}\label{eq_BdGH}
    \mathcal{H}_{\bm{k}}=\frac{S}{2}
 \begin{bmatrix}
  \mathbf{A}(\bm{k}) & \mathbf{B}(\bm{k}) \\
  \mathbf{B}^*(-\bm{k}) & \mathbf{A}^*(-\bm{k})
 \end{bmatrix}.
\end{equation}
Here $\psi _{\bm{k}}=( a_{1,\bm{k}},\cdots ,a_{N,\bm{k}},a^\dagger_{1,-\bm{k}},\cdots ,a^\dagger_{N,-\bm{k}} ) ^T$ represent the Nambu basis. The block of Eq.~\eqref{eq_BdGH} can be expressed by $J^{ab}_{ni,mj}$~\cite{supplemental, Corticelli2022}
\begin{align}\label{eq_AB}
     A_{ij}( \boldsymbol{k})=&\sum_{m,n}{\frac{1}{2}\left( J_{ni,mj}^+-iD_{ni,mj} \right)}e^{i\boldsymbol{k}\cdot \left( \boldsymbol{R}_{mj}-\boldsymbol{R}_{ni} \right)}\notag\\
      &-\delta _{ij}\sum_{mn\ell}{J_{ni,m\ell}^{zz}},
        \\
    B_{ij}( \boldsymbol{k})=&\sum_{m,n}{\frac{1}{2}\left( J_{ni,mj}^-+i\Gamma_{ni,mj} \right)}e^{i\boldsymbol{k}\cdot \left( \boldsymbol{R}_{mj}-\boldsymbol{R}_{ni} \right)},
\end{align}
where $\bm{R}_{ni}$ is the position of the $i$-th spin in the $n$-th unit cell, $J^+_{ni,mj}=J_{ni,mj}^{xx}+J_{ni,mj}^{yy}$, $J^-_{ni,mj}=J_{ni,mj}^{xx}-J_{ni,mj}^{yy}$, $D_{ni,mj}=J_{ni,mj}^{xy}-J_{ni,mj}^{yx}$, and $\Gamma_{ni,mj}=J_{ni,mj}^{xy}+J_{ni,mj}^{yx}$. By summing over the unit cell index, we can further define the following symbols according to Eq.~\eqref{eq_AB}: $A_{ij}=J_{ij}^+-iD_{ij}-\delta_{ij} \sum_{mn\ell} J_{ni,m\ell}^{zz}$ and $B_{ij}=J_{ij}^-+i\Gamma_{ij}$. One can further check that $A(\bm k)=A^\dagger(\bm k)$ and $B(\bm k)=B^T(-\bm k)$. As a result, $\mathcal{H}$ is Hermitian.

Interestingly, the definition of the intermediate quantities $J^{\pm}$, $D$ and $\Gamma$ agrees with the decomposition of the exchange coefficient. With fixed lower indices, $J^{ab}_{ni,mj}$ is a rank-2 tensor, which can then be decomposed into three parts in the $xy$ subspace: the scalar part, also referred to as the isotropic exchange interaction, is represented by $J^+$, the antisymmetric part, also referred to as the DM interaction, is represented by $D$, the traceless symmetric part is represented by $J^-$ and $\Gamma$, which originates from the spin-orbit coupling just as the DM interaction. We shall emphasize there is $\pi/2$ phase difference between $J^+$ and $D$, as well as $J^-$ and $\Gamma$.

Finally, the thermal Hall conductivity of non-interacting magnons is given by~\cite{Matsumoto2011prl,Matsumoto2011prb}
\begin{equation}
    \label{eq_thermalconductivity}
    \kappa _{\mu\nu}=-\frac{k_{B}^{2}T}{\hbar (2\pi)^2}\sum_{n=1}^N{\int_{BZ}{c_2\left[ \rho \left( \varepsilon _{n,\boldsymbol{k}} \right) \right] (\Omega_{\mu\nu}) _{n,\boldsymbol{k}}\mathrm{d}^2k}},
\end{equation}
where $c_2(\rho)=(1+\rho)(\mathrm{log}\frac{1+\rho}{\rho})^2-(\mathrm{log}\rho)^2-2\mathrm{Li}_2(-\rho)$ with $\mathrm{Li}_2(z)$ being the polylogarithm function, $\rho_n$ is Bose distribution function of the $n$-th band, and $\Omega_{n,\bm{k}}$ is the magnon Berry curvature~\cite{Shindou2013}. \red{We ignore the magnon-magnon interaction in the magnetic Hamiltonian which only causes extrinsic contribution to the magnon thermal Hall effect and can be disentangled from the intrinsic one by studying temperature profile.}

\red{Before closing this section, we want to comment on the role of magnetic field. The equilibrium magnetic state often depends on external magnetic field. As for its fluctuation, the effect of magnetic field is more involving. In collinear ferromagnets, the magnetic field does not explicitly enter the magnon Hamiltonian in Eq.~\eqref{eq_bilinearH}. Yet, it can still affects the magnon thermal Hall effect by re-orienting the equilibrium direction of the spin order. For nonuniform spin orders such as noncollinear antiferromagnetism, the magnetic field can enter Eq.~\eqref{eq_bilinearH} through the Zeeman term: $B\sum_{ni}\cos(\theta_i) S_{ni}^z$, where $\theta_i$ is the angle between the magnetic field and the local spin order. It can be put in a bilinear form of $S_{ni}^x$ and $S_{ni}^y$. In this way it can affect the magnon thermal Hall effect but it does not change the role of the symmetric exchange interaction.}

{\it Generalized Onsager's Relation.}---At the heart of the Onsager's relation is the time reversal operation $T$. In the electronic anomalous Hall effect, $T$ can flip both the spin order and the anomalous Hall conductivity. The Onsager's relation then states that the anomalous Hall conductivity is an odd function of the magnetic order~\cite{NagaosaAnomalous,Landau2013}. In the magnonic degree of freedom, however, neither the magnon Hamiltonian in Eq.~\eqref{eq_bilinearH} nor the Berry curvature can be flipped by $T$. Therefore, using $T$ alone cannot offer any useful information about the structure of the magnon thermal Hall conductivity. 

A practical generalization is the effective time reversal operation $\mathcal{T}=TC_{2x}^s$ \cite{Chen2014, Suzuki2017, Mook2019, Zhang2019, Cheng2016} where the rotation only acts on spin. Since $\mathcal{T}$ can flip both the magnon Berry curvature and the DM interaction, it is usually assumed that the DM interaction parameter plays the role of `magnetic order' and is hence indispensable in the magnon thermal Hall effect\cite{Onose2010,Mook2014,Mook2019,Owerre2016,Owerre2016prb,Cao2015,Laurell2018,Kawano2019, Zhuo2021,Li2021}. This analysis has been applied in ferromagnets and coplanar antiferromagnets where a global $C_2$ spin rotation axis exists\cite{Cheng2016,Mook2019,Zhang2019}. 

However, for general periodic spin orders, there is a gauge freedom in choosing the local spin-x direction\cite{supplemental}. At each site in a unit cell, only the spin-z direction is well defined, and any direction perpendicular to the $z$-th direction can work equally well as the $x$ direction. The spin-x axis on different sites can have no connection. This gauge freedom also exists in ferromagnetic and coplanar antiferromagnetic crystals.

Mathematically, the above issue is related to the spin group of the magnetic crystals. In ferromagnets, the magnetic order is subject to the spin-only group $G=\infty 2^\prime$\cite{Liu2022, spin-group2, spin-group3, spin-group4, spin-group5, spin-group6, Chen2025,Corticelli2022}, where $C_2$ rotation axis is perpendicular to the $C_\infty$ axis~(i.e., the direction of the ferromagnetic order). For general spin structures, the spin group obeyed by the magnetic order should at least contain the direct product of the subgroup for each one of the spin in a unit cell
\begin{equation}\label{eq_ftr}
\tilde{G} = \bigotimes_{i=1}^{N} G_i, 
\end{equation}
where $G_i=\infty 2^\prime$ with the axis defined in the local spin frame.
We emphasize that more symmetry operations can appear that connect the spin orders on different sites. The group $G$ is just the common subgroup obeyed by any periodic spin textures. Although $\tilde{G}$ has been derived previously\cite{Chen2025,Corticelli2022}, its implication in magnon transport is still unclear.


Since our goal is to study the Onsager's relation, we shall focus on the effective time reversal operation in this spin group, i.e., $\mathcal{T}_x=TC_{2x}^s$ and $\mathcal{T}_y=TC_{2y}^s$. Although physically equivalent, the arbitrary substitution of one with the other yields distinct constraints on the model parameters.

We start with choosing $\mathcal{T}_x$ identically on each spin in a unit cell, which is the uniform effective time reversal operation. We emphasize that although the mathematical symbol is the same, $\mathcal{T}_x$ on different spins can involve $C_{2x}$ rotations along different axis in the lattice frame.  In the magnon case, the effective time reversal operation $\mathcal{T}_x$ is the spinless-version of the time reversal operation, and can be identified as $\mathcal{T}_x=-K$ where $K$ stands for complex conjugation. We then find that\cite{supplemental} $\mathcal{T}_x\mathcal{H}_{\bm{k}}(J_{ij}^+,J_{ij}^-, D_{ij}, \Gamma_{ij})\mathcal{T}_x^{-1}=\mathcal{H}_{-\bm{k}}^*(J_{ij}^+,J_{ij}^-, D_{ij}, \Gamma_{ij})=\mathcal{H}_{\bm{k}}(J_{ij}^+,J_{ij}^-, -D_{ij}, -\Gamma_{ij})$. The last equality holds by directly evaluating the Hamiltonian at $-\bm k$. By calculating the Berry curvature in the original parameter sets $(J_{ij}^+,J_{ij}^-, D_{ij}, \Gamma_{ij})$ but with the time reversal operator and in the new parameter sets $(J_{ij}^+,J_{ij}^-, -D_{ij}, -\Gamma_{ij})$, respectively. We then obtain the first Onsager's relation\cite{supplemental}:
\begin{equation}
    \label{eq_onsager1}
    \kappa_{\mu\nu}(J_{ij}^\pm, -D_{ij}, -\Gamma_{ij}) = -\kappa_{\mu\nu}(J_{ij}^\pm, D_{ij}, \Gamma_{ij}).
\end{equation}

The effective time reversal symmetry can also be chosen in a non-uniform manner. We classify the spins in a unit cell into two groups, such that the time reversal operation on spins in the first group is $\mathcal{T}_x$ while that on spins in the remaining group is $\mathcal{T}_y$. There are consequently two types of parameters, i.e., those with $i$ and $j$ within the same group, labeled by $J_{ij}^{\pm,intra}$, $D_{ij}^{intra}$, and $\Gamma_{ij}^{intra}$ and those with $i$ and $j$ belonging to different groups, labeled by $J_{ij}^{\pm,inter}$, $D_{ij}^{inter}$, and $\Gamma_{ij}^{inter}$. The effect of $\mathcal{T}$ on the intra-group parameters is the same as in the uniform case, i.e., $J_{ij}^{\pm,intra}\rightarrow J_{ij}^{\pm,intra}$, $D_{ij}^{intra}\rightarrow -D_{ij}^{intra}$, and $\Gamma_{ij}^{intra}\rightarrow -\Gamma_{ij}^{intra}$. In contrast, its effect on the inter-group parameters is opposite: $J_{ij}^{\pm,inter}\rightarrow -J_{ij}^{\pm,inter}$, $D_{ij}^{inter}\rightarrow D_{ij}^{inter}$, and $\Gamma_{ij}^{inter}\rightarrow \Gamma_{ij}^{inter}$. We can then obtain the second Onsager's relation\cite{supplemental}:
\begin{align}\label{eq_onsager2}
        &\kappa_{\mu\nu}(J_{ij}^{\pm, intra}, -D_{ij}^{intra}, -\Gamma_{ij}^{intra}, -J_{ij}^{\pm, inter}, D_{ij}^{inter}, \Gamma_{ij}^{inter})\notag\\
        &=-\kappa_{\mu\nu}(J_{ij}^{\pm, intra}, D_{ij}^{intra}, \Gamma_{ij}^{intra}, J_{ij}^{\pm, inter}, D_{ij}^{inter}, \Gamma_{ij}^{inter})
\end{align}

{\it The role of the symmetric exchange interaction.---}The two Onsager's relations in Eq.~\eqref{eq_onsager1} and \eqref{eq_onsager2} are main results of this work. They reveal complicated dependence of the magnon Berry curvature on different types of exchange couplings. The first Onsager's relation can obviously reproduce the usual understanding that the magnon thermal Hall effect is induced by the DM interaction 
 with vanishing anisotropic parameter $\Gamma_{ij}$. 

 More importantly, Eq.~\eqref{eq_onsager1} suggests that when the DM intereaction is zero, the magnon thermal Hall effect can still emerge, purely due to the symmetric but anisotropic exchange interaction. In this case, $\kappa_{\mu\nu}$ is an odd function of $\Gamma_{ij}$. Equation~\eqref{eq_onsager2} further yields a necessary condition for this case: for the symmetric exchange between two different sites $i$ and $j$, the $J_{ij}^{\pm}$ is also necessary and the resulting magnon thermal Hall effect is also an odd function of $J_{ij}^{\pm}$. In fact, one can further prove that $\Gamma_{ni,mj}$ is equivalent to $J_{ni,mj}^-$, due to a rotation of the local spin frame at either $i$ or $j$ by $\pi/2$.

\begin{figure}[t]
	\includegraphics[width=\columnwidth]{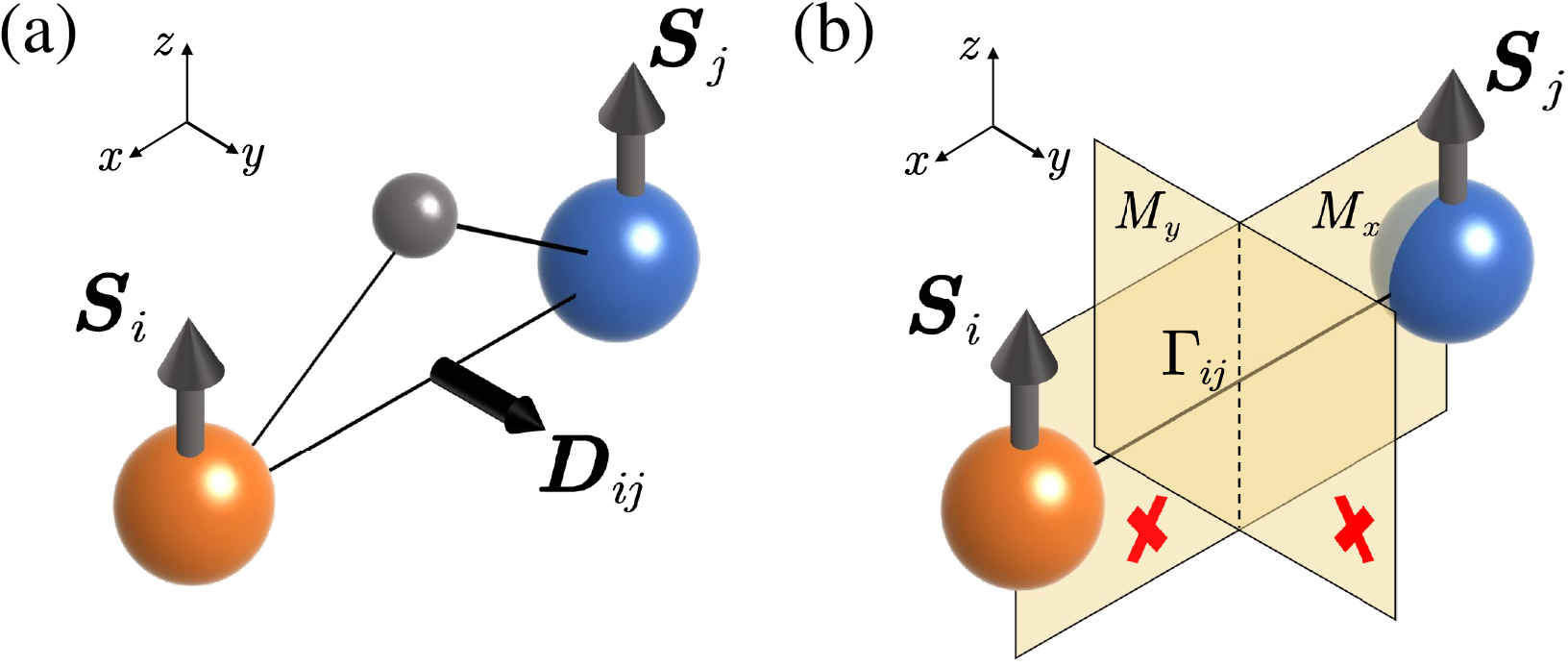}
	\caption{DM interaction (a) and symmetric exchange interaction (b). In (a), the intermediate non-magnetic gray atom  breaks the inversion and mirror-z symmetry. In (b),  the mirror-x and mirror-y symmetry should be broken to have the symmetric coupling $S_{ix}S_{jy}+S_{iy}S_{jx}$. The mirror-z symmetry is allowed.}
	\label{fig1}
\end{figure}

 \red{The symmetric exchange interaction has different symmetry requirements from the DM interaction. The latter requires the breaking of local inversion symmetry. In contrast, to have nonzero symmetric exchange interaction, the local inversion symmetry can exist, and only certain local mirror symmetry needs to be broken, as shown in Fig.\ref{fig1}. For example, to have $J_{xy}(S_{ix}S_{jy}+S_{iy}S_{jx})$, mirror-x and mirror-y symmetry should be broken and mirror-z symmetry can be retained. We shall note that to make this term affect the magnon thermal Hall effect, the equilibrium spin direction should contain perpendicular component to the retained mirror plane.}

 \red{Previously, it is known that besides the DM interaction, the dipole-dipole interaction can also cause a magnon thermal Hall effect\cite{Matsumoto2014, Takeda2024-bv, Gunnink2021}. Here our theory of the symmetric exchange interaction includes the dipole-dipole interaction as a special case. In fact, the latter has the following form $\hat{H}_{dipole}\propto(\hat{e}_{ij}\cdot \bm S_i)(\hat{e}_{ij}\cdot \bm S_j)$, which is always symmetric upon interchanging $\bm S_i$ with $\bm S_j$. However, since the dipole-dipole interaction comes from the electromagnetic force, it is usually much smaller than the one from the electron-electron exchange interaction, provided that the required symmetry breaking for the latter is achieved.} 

\begin{figure}[t]
	\includegraphics[width=\columnwidth]{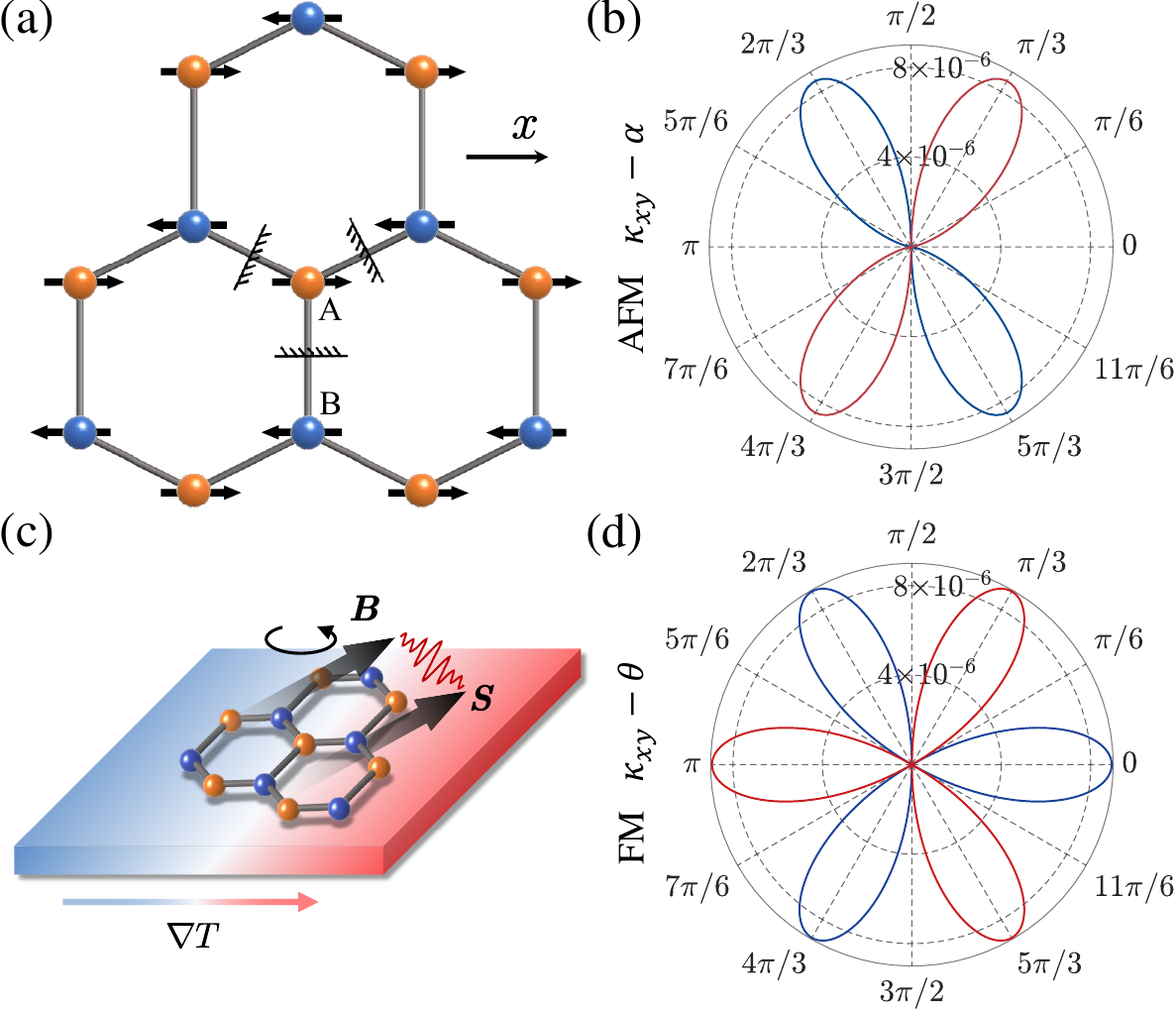}
	\caption{Magnon thermal Hall effect from symmetric exchange interaction. (a) Antiferromagnetic honeycomb lattice, with mirror planes bisecting each nearest neighbor bond. The spin order is along the x direction. (b) The thermal conductivity as a function of parameter $\alpha$ with $\alpha$ being the angle of $J^-+i\Gamma$. The magnitude $\sqrt{(J^-)^2+\Gamma^2}$ is fixed. (c) Schematics of in-plane magnon thermal Hall effect.  (d) Thermal conductivity as a function of the orientation of the in-plane ferromagnetic spin order aligned by external magnetic field.}
	\label{fig:model}
\end{figure}

 \red{To illustrate this type of magnon thermal Hall effect, we consider a honeycomb lattice with in-plane antiferromagnetic order, as shown in Fig.~\ref{fig:model}(a). We choose the honeycomb lattice as there are two magnetic atoms in a unit cell, which is essential for generating a nontrivial magnon Berry curvature. Otherwise, one should consider a sample with finite dimension along one direction to achieve the same purpose. For each nearest neighbor bond, we assume that the inversion symmetry and the mirror symmetry that interchanges neighboring magnetic atoms are retained. Therefore, the nearest neighbor DM interaction vanishes. The Hamiltonian is given by 
  \begin{equation}\label{eq_ham}
    H = \sum_{\langle i,j\rangle}{\boldsymbol{S}_{i}^T\mathcal{J} _{ij}\boldsymbol{S}_j} -\sum_i\bm B\cdot \bm S_i.
\end{equation}
The first term takes account the isotropic and anisotropic symmetric exchange coupling, consistent with the symmetry, and the second term is the Zeeman energy from external magnetic field along the $x$-direction, which breaks the chiral symmetry $\sigma_x K$ and hence induces a finite band gap for a nontrivial Berry curvature~\cite{supplemental}. The detail of parameters can be found in Ref.~\cite{supplemental}.}

\red{In Fig.~\ref{fig:model}(b) we plot the thermal Hall conductivity as a function of $\Gamma$ and $J^-$. One can find a $d$-wave pattern, showing that $\kappa_{xy}$ is an odd function of both $\Gamma$ and $J^-$, consistent with the two Onsager's relation in Eq.~\eqref{eq_onsager1} and \eqref{eq_onsager2}. It demonstrates the necessary condition for this type of thermal Hall conductivity, i.e., a nonzero $J^-$ is also required.}

 \red{In the electronic anomalous Hall effect, it is often assumed that the magnetization direction is perpendicular to the Hall-deflection plane due to a widely used empirical law. However, in recent years, it has been found that if the crystal symmetry is sufficiently low, the magnetization can lie within the Hall-deflection plane, causing an in-plane anomalous Hall effect~\cite{Liu2007, Roman2009, Tan2021, Cao2023, Zhou2022}.} 

\red{Such in-plane geometry can persist in the magnonic degree of freedom with the help of the anisotropic exchange interaction. To show this, we consider the same honeycomb lattice as above but with in-plane ferromagnetic spin order. The spin order can then be aligned by external magnetic field, as shown in Fig.~\ref{fig:model}(c). The magnon Hamiltonian is similar with Eq.~\eqref{eq_ham}. An additional staggered isotropic exchange term between second nearest neighbors is added to break the chiral symmetry, which can be achieved by strain gradient, Zeeman field gradient, in-plane electric field, etc..}

\red{We then calculate the thermal Hall conductivity as a function of the orientation of the in-plane spin order. As shown in Fig.~\ref{fig:model}(d), the result  is nonzero and exhibits a three-fold rotational symmetry,  consistent with the $C_3$ rotational symmetry. Moreover, when the rotation angle is $\pi$, i.e., the spin order is flipped, the spin frame for magnons also change by a $C_{2x}$ rotation which reverses the sign of $\Gamma$ without changing other properties of the magnon spectrum. According to Onsager's relation in Eq.~\eqref{eq_onsager1}, thermal conductivity should also flip sign, as shown in Fig.~\ref{fig:model}(d). This exotic pattern is analogous to those discovered in electronic anomalous Hall effect~\cite{Wang2024,Battilomo2021,Wang2023,Peng2024}.}

\red{Symmetrywise speaking, the generation of the in-plane magnon thermal Hall effect is similar to that of the in-plane anomalous Hall effect, i.e., there should be no $C_2$ axis perpendicular to the Hall-deflection plane\cite{Liu2024}. The only out-of-plane rotation axis allowed is thus $C_3$. In-plane $C_2$ axis is also allowed, and when the spin order aligns with this axis, the thermal conductivity vanishes. This occurs in materials with point group $D_3$, $D_{3d}$.} 

\red{We shall emphasize that the symmetry analysis is only for a specific Hall-deflection plane. As for the original crystals, the point group symmetry can be quite high. For example, for materials with cubic point group such as $T$, $T_d$, $T_h$, $O$, $O_h$, the in-plane magnon thermal Hall effect can still be discovered as long as the Hall deflection plane is properly selected to fulfill the above symmetry requirement, such as the $(112)$-plane.}

 \red{The mechanism from the symmetric but anisotropic exchange interaction  expands the potential systems with magnon thermal Hall effect, in the way that it allows re-examination of magnetic crystals that preserves local inversion symmetry or that the local DM interaction contains a $\bm D$ vector perpendicular to the spin order. One promising candidate is ferromagnetic material $\mathrm{VAu_4}$. The middle point of the nearest V atoms is an inversion center\cite{supplemental} so that DM interaction is forbidden. Moreover, the two mirror symmetries shown in FIG.\ref{fig:model}(b) are broken and hence the symmetric exchange interaction is allowed which can cause a nonzero magnon thermal Hall effect in a thin-film $\mathrm{VAu_4}$. Similar systems include layered $\mathrm{CrCl_3}$ and $\mathrm{V_2Se_2O}$ with in-plane electric field or strain gradient.}

\red{In summary, we establish two generalized Onsager's relations in magnon thermal Hall effect from which we propose a distinct mechanism of the magnon thermal Hall effect solely from the symmetric but anisotropic exchange interaction. The mechanism can compete with the previously proposed mechanisms from the DM interaction and dipole-dipole interaction. More importantly, as it involves different types of symmetry breaking, it can expands systems with the magnon thermal Hall effect. As an application, we show that it can leads to a special in-plane magnon thermal Hall effect.}

\begin{acknowledgments}
We acknowledge useful discussions with Zheng Liu, Shiyue Deng, Shu Li and Qingtao Zhao. This work is supported by the National Natural Science Foundation of China (12234017). Y. G. is also supported by the Innovation Program for Quantum Science and Technology (2021ZD0302802). The supercomputing service of USTC is gratefully acknowledged.
\end{acknowledgments}

\end{document}